\documentclass[twocolumn,preprintnumbers,amsfonts,amsmath,amssymb,nofootinbib,floatfix]{revtex4-1}

\usepackage{epsfig}
\usepackage{latexsym}
\usepackage{graphicx}
\usepackage{dcolumn}
\usepackage{bm}

\newcommand{\be}{\begin{equation}}
\newcommand{\ee}{\end{equation}}
\newcommand{\ba}{\begin{eqnarray}}
\newcommand{\ea}{\end{eqnarray}}
\newcommand{\bi}{\begin{itemize}}
\newcommand{\ei}{\end{itemize}}

\newcommand{\msbar}{{\rm \overline{MS\kern-0.05em}\kern0.05em}}

\newcommand{\wS}{\ensuremath{\widetilde\rho} }

\graphicspath{{figs/}}

\begin{document}

\title{Chiral symmetry breaking in QCD Lite}

\author{Georg P.~Engel$^{a}$, Leonardo Giusti$^{a}$, Stefano Lottini$^{b}$,
Rainer Sommer$^{b}$}

\affiliation{\vspace{0.3cm}
\vspace{0.1cm} 
$^a$ Dipartimento di Fisica, Universit\`a di Milano-Bicocca, and 
INFN, Sezione di Milano-Bicocca, Piazza della Scienza 3, 
I-20126 Milano, Italy\\
$^b$ John von Neumann Institute for Computing (NIC), DESY, Platanenallee 6, D-15738 Zeuthen, Germany}

\date{\vspace{0.2cm} \today}

\begin{abstract}
A distinctive feature of the presence of spontaneous chiral symmetry breaking in QCD
is the condensation of low modes of the Dirac operator near the origin.
The rate of condensation must be equal to the slope of $M_\pi^2 F_\pi^2/2$ 
with respect to the quark mass $m$ in the chiral limit, where $M_\pi$ and 
$F_\pi$ are the mass and the decay constant of the Nambu-Goldstone bosons. 
We compute the spectral density of the (Hermitian) Dirac operator, the quark mass,
the pseudoscalar meson mass and decay constant by numerical simulations 
of lattice QCD with two light degenerate Wilson quarks. We use CLS lattices
at three values of the lattice spacing in the range $0.05$--$0.08$~fm, and for 
several quark masses corresponding to pseudoscalar mesons masses down to $190$~MeV. 
Thanks to this coverage of parameters space, we can extrapolate all quantities to 
the chiral and continuum limits with confidence. The results show that the low quark 
modes do condense in the continuum as expected by the Banks--Casher mechanism, and the 
rate of condensation agrees with the Gell--Mann-–Oakes-–Renner (GMOR) relation. For the 
renormalisation-group-invariant ratios we obtain 
$[\Sigma^{\rm RGI}]^{1/3}/F =2.77(2)(4)$ 
and $\Lambda^{\msbar}/F  =  3.6(2)$, which correspond to 
$[\Sigma^\msbar(2\, \mbox{GeV})]^{1/3} =263(3)(4)$~MeV and $F=85.8(7)(20)$~MeV 
if $F_K$ is used to set the scale by supplementing the 
theory with a quenched strange quark.
\end{abstract}

\maketitle

\noindent {\it Introduction.---}  
There is overwhelming evidence that the chiral symmetry 
group $SU(N_f)_L\times SU(N_f)_R$ of Quantum Chromodynamics 
(QCD) with a small number $N_f$ of light flavours 
breaks spontaneously to $SU(N_f)_{L+R}$. 
This progress became possible over the last 
decade thanks to the impressive speed-up of the numerical 
simulations of lattice QCD with light dynamical fermions, 
see Ref.~\cite{Schaefer:2012tq} for a recent review and a 
comprehensive list of references. The impact on phenomenological 
analyses of chiral dynamics is already striking~\cite{Leutwyler:2008ma}.

The formation of a non-zero chiral condensate in the theory, 
$\Sigma=-\frac{1}{2}\langle\bar\psi \psi\rangle|_{m=0}$, 
was conjectured to be the effect of the  condensation of 
the low modes of the Dirac operator near the 
origin~\cite{Banks:1979yr}. 
The rate of condensation is indeed a renormalisable 
universal quantity in QCD, and is unambiguously defined 
once the bare parameters in the action of the theory have 
been renormalised~\cite{Giusti:2008vb}. The Banks--Casher
mechanism links the spectral density $\rho(\lambda,m)$ 
of the Dirac operator to the condensate as \cite{Banks:1979yr}
\be\label{eq:BK}
   \lim_{\lambda \to 0}\lim_{m \to 0}\lim_{V \to \infty}\rho(\lambda,m)
   =\frac{\Sigma}{\pi}\; ,
\ee
an identity which can be read in both directions: a non-zero 
spectral density implies that the symmetry is broken by a 
non-vanishing $\Sigma$ and vice versa. 

The above conceptual and technical advances in lattice gauge theory paved 
the way for a quantitative study of the Banks--Casher mechanism from 
first principles. It is the aim of this letter to achieve a precise and 
reliable determination of the density of eigenvalues
$i\lambda$ of the Euclidean Dirac operator $D$ near the origin at small 
quark masses in the continuum. As in any numerical computation,
the limits in Eq.~(\ref{eq:BK}) inevitably require an extrapolation of the 
results with a pre-defined functional form. The distinctive signature for spontaneous
symmetry breaking is the agreement between the chiral-limit value 
of the spectral density at the origin, reached by extrapolating the data 
with the functional form dictated by chiral perturbation theory (ChPT), 
and the slope of $M_\pi^2 F_\pi^2/2$ with respect to the 
quark mass $m$ \cite{Weinberg:1978kz,Gasser:1984gg}. We thus complement 
our study with the computations of $m$, $M_\pi$ and $F_\pi$.

To reach these goals, we use $O(a)$-improved Wilson fermions at 
several lattice spacings, and we extrapolate the numerical results 
to the universal continuum limit following the Symanzik effective 
theory analysis. For technical reasons we focus on the mode number 
of the Dirac operator~\cite{Giusti:2008vb}
\vspace{-0.625cm}

\be
  \nu(\Lambda,m)=V \int_{-\Lambda}^{\Lambda} {\rm d}\lambda\,\rho(\lambda,m),
\ee
which at the same time is the average number of eigenmodes of the massive 
Hermitian operator 
$D^{\dagger}D+m^2$ with eigenvalues $\alpha\leq M^2 = \Lambda^2+m^2$.
\begin{figure*}[!t]
\begin{center}
\hspace{-1.5cm}\begin{minipage}{0.35\textwidth}
\includegraphics[width=6.5 cm,angle=0]{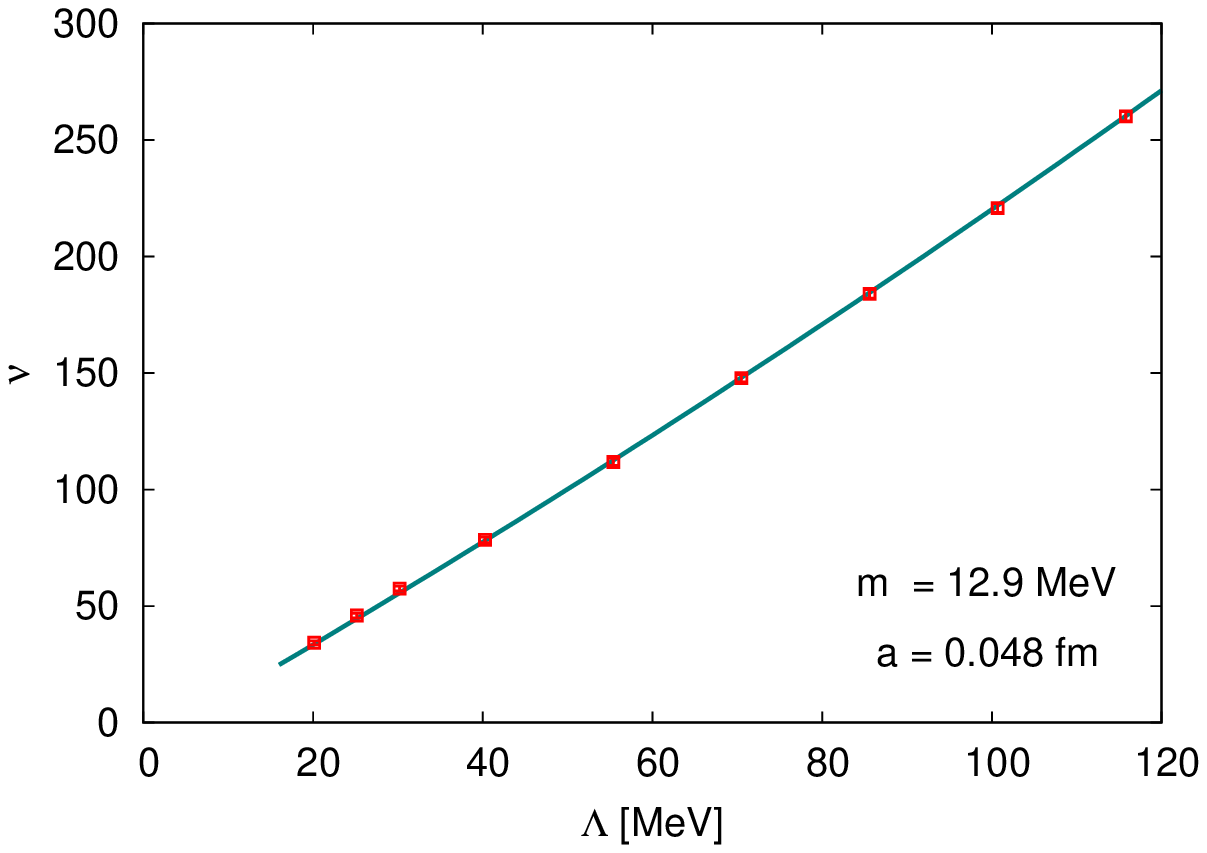}
\end{minipage}
\hspace{1.5cm}
\begin{minipage}{0.35\textwidth}
\includegraphics[width=6.5 cm,angle=0]{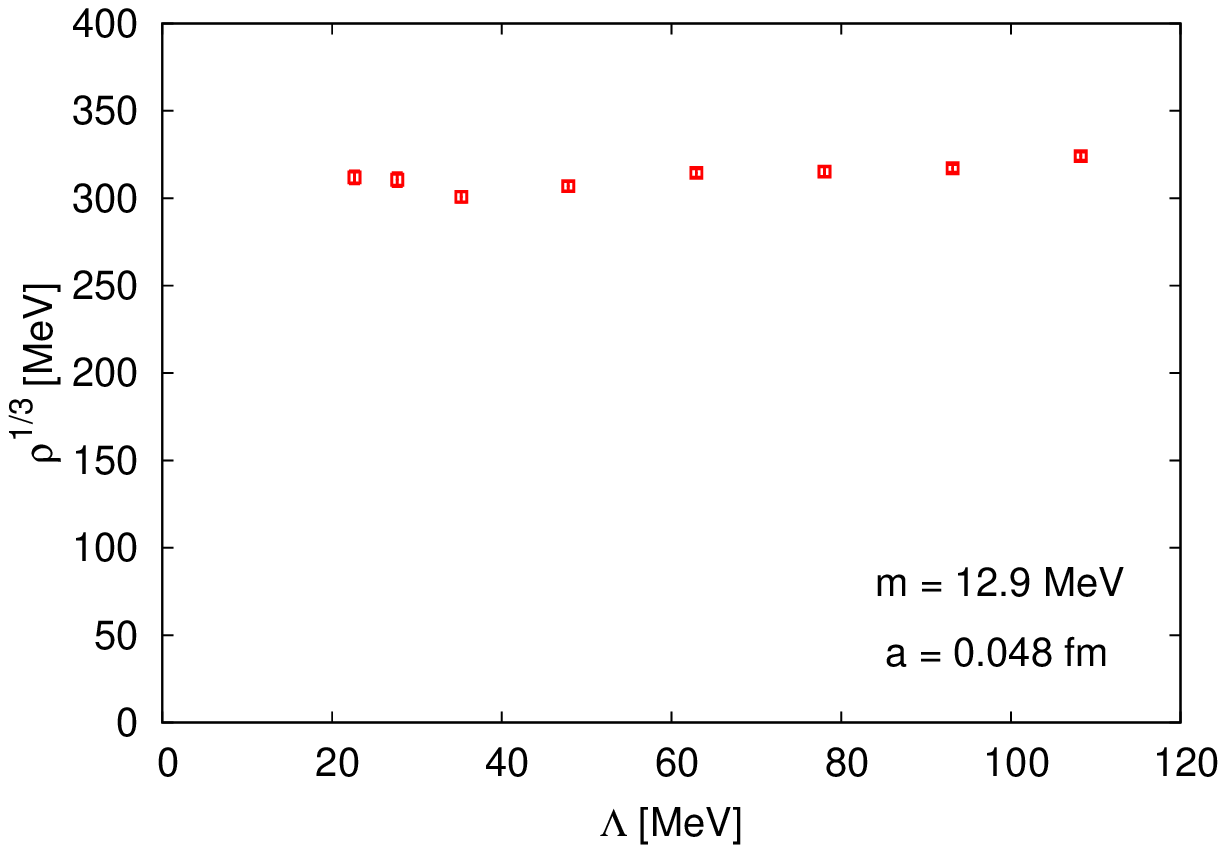}
\end{minipage}

\hspace{-1.5cm}\begin{minipage}{0.35\textwidth}
\includegraphics[width=6.5 cm,angle=0]{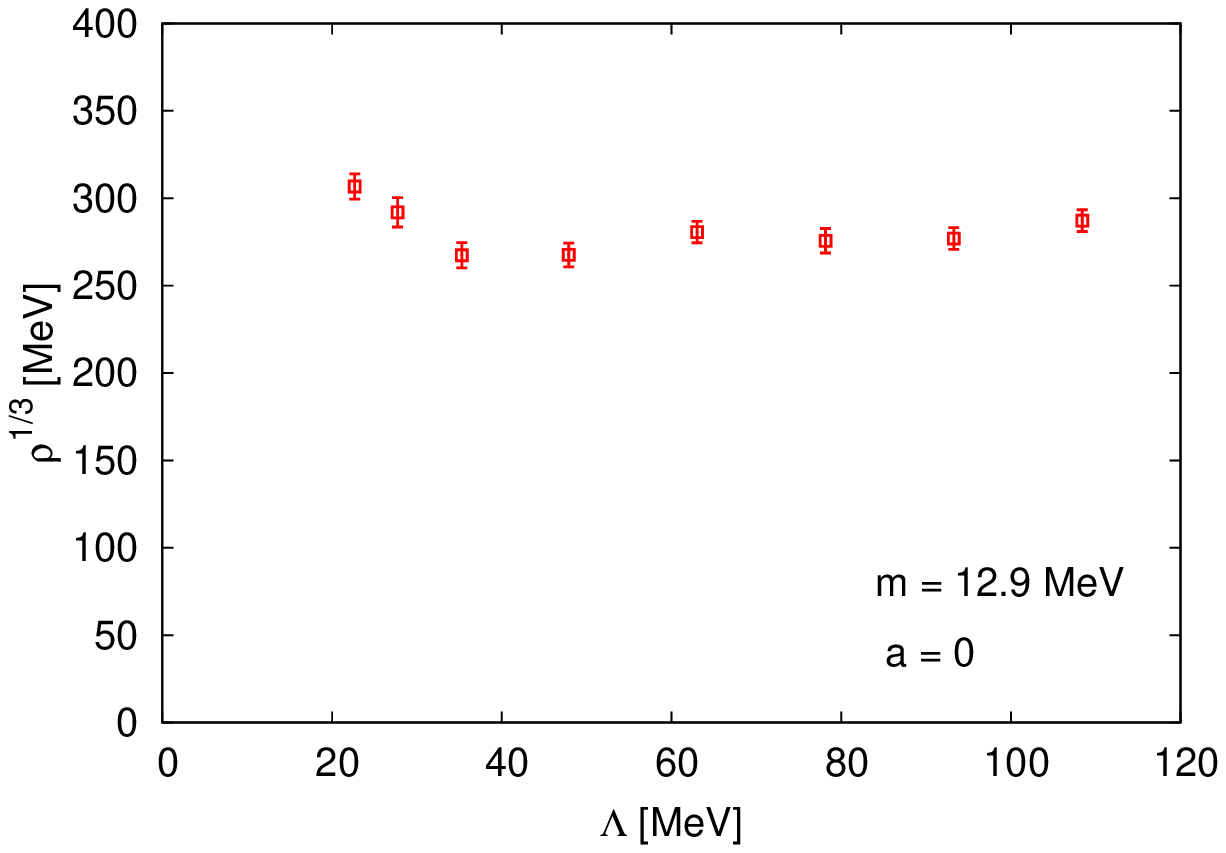}
\end{minipage}
\hspace{1.5cm}
\begin{minipage}{0.35\textwidth}
\includegraphics[width=6.5 cm,angle=0]{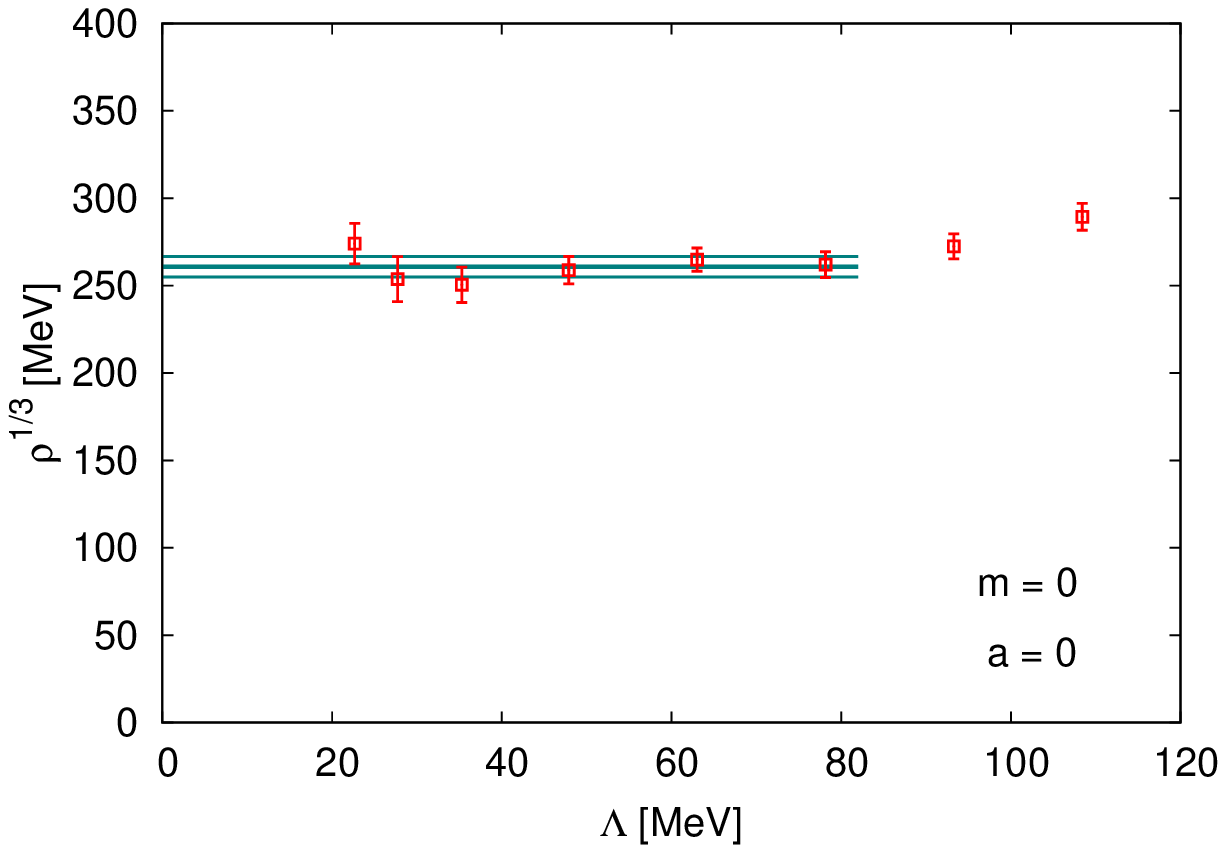}
\end{minipage}
\end{center}
\vspace{-0.50cm}

\caption{Top-left: the mode number as a function of $\Lambda$ for O7; a quadratic fit of the data ($\Lambda$ in MeV) 
gives $\nu= -9.0(13) + 2.07(7)\Lambda + 0.0022(4)\Lambda^2$. Top-right: $\wS^{\, 1/3}$ for the 
same ensemble as a function of $\Lambda=(\Lambda_1+\Lambda_2)/2$. Bottom-left:
$\wS^{\, 1/3}$ in the continuum limit at the smallest reference quark mass. 
Bottom-right: $\wS^{\, 1/3}$ in the continuum and chiral limit. Note the flat dependence on 
$\Lambda$ which agrees with the expectation from NLO ChPT: a plateau fit in the interval 
20--80 MeV is also shown.}
\label{fig:firstlook-O7}
\vspace{-0.5cm}

\end{figure*}
It is a renormalisation-group-invariant quantity as it stands. Its 
(normalised) discrete derivative 
\be\label{eq:discD}
{\widetilde \rho}(\Lambda_1,\Lambda_2,m)  =  
\frac{\pi}{2 V} \frac{\nu(\Lambda_2)-\nu(\Lambda_1)}
{\Lambda_2 - \Lambda_1}\;  
\ee
carries the same information as $\rho(\lambda,m)$, but the {\it effective
density} ${\widetilde \rho}$ is a more convenient quantity to consider 
in numerical computations. For practical 
purposes we also extend the theory by introducing a quenched ``strange'' 
quark so to have a graded chiral symmetry group $SU(3|1)$. This is instrumental 
to derive the ChPT formula for $\rho(\lambda,m)$~\cite{Osborn:1998qb}, and 
allows us to fix the lattice spacing from the kaon decay constant 
$F_K$. The latter is a well-defined intermediate reference 
scale which can be computed precisely on the lattice~\cite{Fritzsch:2012wq} and is 
directly accessible to experiments once the CKM matrix element $|V_{us}|$ is known. 
This scale is used here to convert all quantities in physical units, with 
the scheme-dependent ones renormalised in the $\msbar$ scheme at 
$\mu=2$~GeV. The final results, however, 
are independent of this intermediate step: they are expressed as ratios of quantities 
of the two-flavour theory only.\\ 
\indent It is worth noting that there were several exploratory 
studies of the spectral density of the Dirac operator in QCD, 
see for instance~\cite{Giusti:2008vb,Fukaya:2010na,Cichy:2013gja}. 
The approach pursued here is rather general, and it may be 
useful in order to study theories at non-zero temperature or 
strongly interacting models of electroweak symmetry 
breaking~\cite{Patella:2012da}.

\noindent {\it Lattice computation.---} 
We have profited from CLS simulations of 
two-flavour QCD with the $O(a)$-improved Wilson 
action. On all the lattices listed in Table~\ref{tab:ens}
we have computed the mode number and the two-point functions of 
$\bar\psi_1\gamma_5\psi_2$ and $\bar\psi_1\gamma_0\gamma_5\psi_2$.
The ensembles have lattice spacings of $\;a=0.075,\, 0.065,\, 0.048$\,fm
as measured from $F_K$ \cite{Fritzsch:2012wq}. The quark masses range 
from $6$ to $40$~MeV and are small compared to the typical scale of the 
theory from the condensate or the string tension of about $250$-$450$~MeV.
All lattices are of size $2L \times L^3$, and the pion mass is always 
large enough so that $m_\pi L\geq4$. Finite-size effects are within the 
statistical errors for all measured quantities, see Ref.~\cite{Engel:2014prep} 
for more details. The error analysis takes care of autocorrelations~\cite{Schaefer:2010hu}, 
all Markov chains except for one (N5) being of length between $24$ and 74 $\tau_{\rm exp}$.\\ 
\indent The mode number has been computed for nine values of $\Lambda$ in the range 
$20$--$120$~MeV with a statistical accuracy of a few percent on all lattices. 
Four larger values of $\Lambda$ in the range $150$--$500$~MeV have also been 
analyzed for the ensemble E5, see Ref.~\cite{Engel:2014prep} for tables with
all results. In Fig.~\ref{fig:firstlook-O7} (top-left) we show $\nu$ as a 
function of $\Lambda$ for the lattice O7, corresponding to the smallest quark 
mass at the smallest lattice spacing. On all other lattices an analogous 
qualitative behaviour is observed. The mode number is a nearly linear function 
in $\Lambda$ up to approximately $100$--$150$~MeV. A clear departure from 
linearity is observed for $\Lambda > 200$~MeV on the lattice E5.  At the percent 
precision, however, the data show statistically significant deviations from 
the linear behavior already below $100$ MeV. To guide the eye, a quadratic fit 
in $\Lambda$ is shown in Fig.~\ref{fig:firstlook-O7}, and the values of the 
coefficients are given in the caption. The bulk of $\nu$ is given by the linear 
term, while the constant and the quadratic term represent $O(10\%)$ corrections 
in the fitted range. The nearly linear behaviour of the mode number, expected
if the Banks--Casher mechanism is at work, is manifest 
on the top-right plot of Fig.~\ref{fig:firstlook-O7}, where the cubic root 
of the discrete derivative defined in Eq.~(\ref{eq:discD}) is shown
as a function of $\Lambda =(\Lambda_1+\Lambda_2)/2$ for each couple of 
consecutive values of $\Lambda$.
When the regularisation breaks chiral symmetry, discretization effects heavily 
distort the spectral density near $\lambda=0$~\cite{DelDebbio:2005qa,Damgaard:2010cz}: 
we thus focus on the effective spectral density rather than the mode number.\\
\begin{table}[!t]
\small
\begin{center}
\setlength{\tabcolsep}{.3pc}
\begin{tabular}{@{\extracolsep{0.0cm}}ccccccccc}
\hline
id &$L/a$&$m$~[MeV]&$F_\pi$~[MeV]&$M_\pi$~[MeV]&$M_\pi L$&$a$~[fm]\\
\hline
A3  &$32$&$37.4(9)$& $120.8(7)$ &$496(6)$ & $6.0$ & 0.0749(8)\\     
A4  &$32$&$22.8(6)$& $110.7(6)$ &$386(5)$ & $4.7$  & \\
A5  &$32$&$16.8(4)$& $106.0(6)$ &$333(5)$ & $4.0$  & \\
B6  &$48$&$12.2(3)$& $102.3(5)$ &$283(4)$ & $5.2$  & \\
\hline
E5  &$32$&$32.0(8)$& $115.2(6)$ &$440(5)$ & $4.7$ & 0.0652(6)\\
F6  &$48$&$16.5(4)$& $105.3(6)$ &$314(3)$ & $5.0$ & \\
F7  &$48$&$12.0(3)$& $100.9(4)$ &$268(3)$ & $4.3$ & \\
G8  &$64$&$\;\;6.1(2)$&$\;\;95.8(4)$  &$193(2)$ & $4.1$ & \\
\hline
N5  &$48$&$34.8(8)$& $115.1(7)$ &$443(4)$ & $5.2$ & 0.0483(4)\\
N6  &$48$&$20.9(5)$& $105.8(5)$ &$342(3)$ & $4.0$ & \\
O7  &$64$&$12.9(3)$& $101.2(4)$ &$269(3)$ & $4.2$ & \\
\hline
\end{tabular}
\end{center}
\vspace{-0.25cm}

\caption{Overview of the ensembles used in this
study. We give label, spatial extent of the lattice,
quark mass $m$, pion mass $M_\pi$ and its decay constant
$F_\pi$,  and the (updated) value of the lattice spacing 
determined from $F_K$ as in Ref.~\cite{Fritzsch:2012wq}.}
\label{tab:ens}
\vspace{-0.55cm}

\end{table}
\indent In general, $\wS^{\, 1/3}$ shows quite a flat behaviour in 
$\Lambda$ at fine lattice spacings and light quark masses, similar to the 
one shown in Fig.~\ref{fig:firstlook-O7} (top-right). Because the action and 
the mode number are $O(a)$-improved, the Symanzik effective-theory analysis
predicts that discretization errors start at $O(a^2)$~\cite{Giusti:2008vb}. 
In order to remove them, we interpolate the effective spectral density to 
three quark mass values ($m=12.9$, $20.9$, $32.0$~MeV) at each lattice spacing. 
The values of $\wS^{\,1/3}$ show very mild discretization effects at light $m$
and $\Lambda$, while they differ by up to $15\,\%$ among the three 
lattice spacings toward heavier $\Lambda$. Within the statistical 
errors all data sets are compatible with a linear dependence in $a^2$, and 
we thus independently extrapolate each triplet of points to the continuum 
limit accordingly. The difference between the values of $\wS^{\,1/3}$ at the 
finest lattice spacing and the continuum-extrapolated ones is within 
$5\,\%$ for the lightest $m$ and $\Lambda$, and it remains within few 
standard deviations at 
heavier values of $m$ and $\Lambda$. This makes us confident that the 
extrapolation removes the cutoff effects within the errors quoted.\\
\indent The results for $\wS^{\, 1/3}$ at $m=12.9$~MeV in the continuum limit are shown 
as a function of $\Lambda$ in the bottom-left plot of Fig.~\ref{fig:firstlook-O7}. 
A similar $\Lambda$-dependence is observed at the two other reference masses. 
It is noteworthy that no assumption on the presence of spontaneous symmetry 
breaking was needed so far. These results point to the fact that 
the spectral density of the Dirac operator in two-flavour QCD is non-zero 
and (almost) constant in $\Lambda$ near the origin at small quark masses. This is 
consistent with the expectations from the Banks--Casher mechanism. 
In presence of spontaneous symmetry breaking, next-to-leading (NLO) ChPT indeed 
predicts~\cite{Leutwyler:1992yt,Smilga:1993in,Osborn:1998qb,Giusti:2008vb}
\ba\label{eq:eqChPTtext}
\wS^{\rm\; nlo} & = &    
\Sigma \Big\{ 1  + \frac{m \Sigma}{(4\pi)^2 F^4} \Big[3\, \bar l_6 + 1 - 
\ln(2) \nonumber\\ 
& - & 3 \ln\Big(\frac{\Sigma m}{F^2 \bar\mu^2}\Big) + 
\tilde g_\nu\left(\frac{\Lambda_1}{m},\frac{\Lambda_2}{m}\right) 
\Big]\Big\}\; ,
\ea
i.e.~an almost flat function in (small) $\Lambda$ at (small) finite quark 
masses.\footnote{The parameter $\bar l_6$ is a low-energy constant of 
the $SU(3|1)$ chiral effective theory renormalised at the scale $\bar\mu$,
while $\tilde g_\nu$ is a parameter-free function, see Ref.~\cite{Engel:2014prep}.} 
Once the pion mass and decay constant 
are measured, the (mild) parameter-free $\Lambda$-dependence of 
$\wS^{\rm\; nlo}$ in Eq.~(\ref{eq:eqChPTtext}) is compatible 
with our data.\\ 
\indent The extrapolation to the chiral limit requires an assumption 
on how the effective spectral density $\wS$ behaves when 
$m \rightarrow 0$. In this respect it is interesting to notice that
toward the chiral limit the function in Eq.~(\ref{eq:eqChPTtext}) is 
the simplest possible one, i.e.~it goes linearly in $m$ since there are no 
chiral logarithms at fixed $\Lambda$~\cite{Giusti:2008vb}. A fit of the data 
to Eq.~(\ref{eq:eqChPTtext}) shows that they are compatible with 
that NLO formula. Eq.~(\ref{eq:eqChPTtext})
predicts that in the chiral limit $\wS^{\rm\; nlo}=\Sigma$ also at non-zero $\Lambda$, 
since all NLO corrections vanish~\cite{Smilga:1993in}. By extrapolating the
effective spectral density with 
Eq.~(\ref{eq:eqChPTtext}) but allowing for the constant term to depend 
on $\Lambda$, we obtain the results shown in the bottom-right plot of 
Fig.~\ref{fig:firstlook-O7}  with a $\chi^2/\rm{dof}=16.4/14$.
Within errors the $\Lambda$-dependence is clearly compatible with 
a constant up to $\approx 100$~MeV. Moreover the differences between 
the values of $\wS^{\, 1/3}$ in the chiral limit and those at $m=12.9$~MeV 
are of the order of the statistical error, i.e.~the extrapolation is 
very mild. A fit to a constant of the data gives 
$\Sigma^{1/3} = 261(6)(8)$~MeV, 
where the first error is statistical and the 
second one is systematic. The latter is a conservative estimate
obtained by performing various combined fits of all data 
suggested by NLO ChPT and the Symanzik effective theory 
analysis~\cite{Engel:2014prep}.\\
\begin{figure}[!t]
\begin{center}
\includegraphics[width=7.5 cm,angle=0]{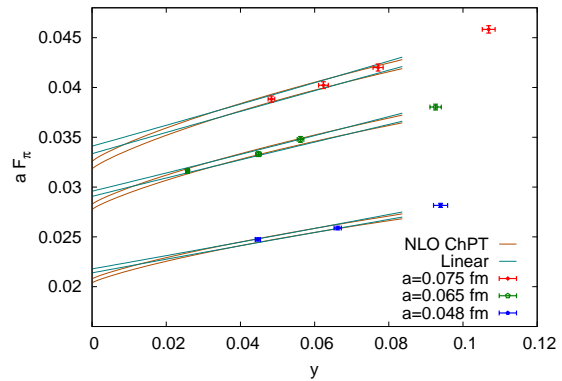}
\vspace{-0.25cm}

\caption{The pseudoscalar decay constant $a F_\pi$ versus 
\mbox{$y=M_\pi^2/(4 \pi F_\pi)^2$}. The bands are the result of a combined fit, 
see main text.}
\label{fig:Ffinal}
\end{center}
\vspace{-.75cm}

\end{figure}
\indent To compare the value of the spectral density at the origin with the slope of 
$M_\pi^2 F_\pi^2/2$ with respect to the quark mass $m$, we complement 
the computation of the mode number with those for the pion masses 
and the decay constants, 
$M_\pi$ and $F_\pi$, as well as the quark mass $m$. They are extracted 
from the two-point functions of the non-singlet pseudoscalar density 
and axial current as in Refs.~\cite{DelDebbio:2007pz,Fritzsch:2012wq}, 
see Ref.~\cite{Engel:2014prep} for more details. The results are 
reported in Table~\ref{tab:ens}, and those for the pseudoscalar decay 
constant in lattice units are shown in Fig.~\ref{fig:Ffinal} versus 
$y=M_\pi^2/(4 \pi F_\pi)^2$. We fit $F_\pi$ to the function
\be
a F_\pi = (a F)\, \{1 - y \ln(y) + b y\}\; ,
\ee 
where $b$ is common to all lattice spacings,
restricted to the points with $M_\pi<\, 400$~MeV 
(see Fig.~\ref{fig:Ffinal}). 
Apart for the NLO ChPT just given, we also perform a number of 
alternative extrapolations in $y$. As a final result we quote 
$a F=0.0330(4)(8)$, $0.0287(3)(7)$ and $0.0211(2)(5)$ at 
$a=0.075$, $0.065$ and $0.048$~fm respectively, where the second (systematic) 
error takes into account the spread of the results from the various fits. 
By performing a continuum-limit extrapolation we obtain our final 
result $F=85.8(7)(20)$~MeV.
\begin{figure}[!t]
\begin{center}
\includegraphics[width=8.0 cm,angle=0]{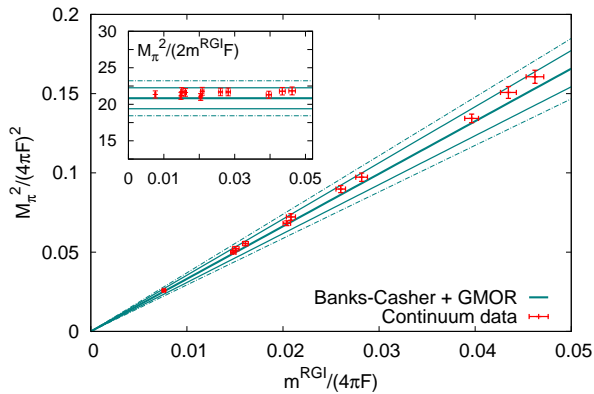}
\vspace{-0.50cm}
\caption{The pion mass squared versus the RGI quark mass, both  
normalised to $4 \pi F$ which is roughly $1$~GeV. The ratio 
$(M_\pi^2/2m F)^{1/3}$ is extrapolated to the continuum as in 
Eq.~(\ref{eq:Mpicont}). The central line is the GMOR 
contribution to the pion mass squared computed by taking the 
direct measure of the condensate from the spectral density. 
The upper and lower solid lines show 
the statistical error and the dotted-dashed ones the total error, 
the systematic being added in quadrature.} 
\label{fig:Mpi}
\end{center}
\vspace{-0.825cm}

\end{figure}
Once the value of $F$ is determined, we compute the ratio $M_\pi^2/2m $ 
for all data points. We fit the data with $M_\pi<\, 400$~MeV to 
\be\label{eq:Mpicont}
\Big[\frac{M_\pi^2}{2m F}\Big]^{1/3}=(s_0+s_1 (aF)^2)\{1+ \frac{y}{6} \ln(y) +d\, y\}\; , 
\ee
where again $d$ is common to all lattice spacings. 
Also in this case we checked several 
variants although the data look very flat up to the heaviest mass. 
The result for the condensate is 
$[\Sigma^\msbar_{\rm GMOR}(2\, \mbox{GeV})]^{1/3}= 263(3)(4)$~MeV, where
the errors are determined as for $F$. 

\noindent {\it Discussion and conclusions.---} 
From the previous analysis, our best results for the 
leading-order low-energy constants of QCD with two flavours are
\ba
& & [\Sigma^\msbar(2\, \mbox{GeV})]^{1/3} =  263(3)(4)\; {\rm MeV}\; ,\nonumber\\[0.25cm] 
& & \qquad\;\qquad\qquad F=85.8(7)(20)\; {\rm MeV}\; .
\ea
By updating the value of the $\Lambda$-parameter in 
Ref.~\cite{DellaMorte:2004bc,Fritzsch:2012wq}
to $\Lambda^{\msbar}=311(19)$~MeV and by taking into account the correlation
with $F$, we obtain the dimensionless ratios
\be
\frac{[\Sigma^{\rm RGI}]^{1/3}}{F} =  2.77(2)(4)\;,\quad
\frac{\Lambda^{\msbar}}{F}  =  3.6(2)\; . 
\ee
where the renormalisation-group-invariant (RGI) condensate 
is defined with the convention 
of Refs.~\cite{Capitani:1998mq,DellaMorte:2005kg}.\\
\indent Our results show that the spectral density of the Dirac operator in 
the continuum is non-zero at the origin and that its value agrees
with the slope of $M_\pi^2 F_\pi^2/2$ with respect to the quark mass
when both are extrapolated to the chiral limit. If expanded in $m$, 
$M_\pi^2$ is dominated by the leading (GMOR) term 
proportional to the chiral condensate, see Fig.~\ref{fig:Mpi}. 
The ratio $M_\pi^2/2m$ is nearly constant within errors up to 
quark masses that are about one order of magnitude larger 
than in Nature.

Measurements have been performed on BlueGene/Q at CINECA 
(CINECA-INFN agreement, ISCRA project IsB08\_Condnf2), on HLRN, on JUROPA/JUQUEEN at 
J\"ulich JSC, on PAX at DESY, Zeuthen, and on Wilson at Milano--Bicocca. 
We thank these institutions for the computer resources and the technical 
support. We are grateful to our colleagues within the CLS initiative for 
sharing the ensembles of gauge configurations. G.P.E.~and L.G.~acknowledge 
partial support by the MIUR-PRIN contract 20093BMNNPR. 
S.L.~and R.S.~acknowledge support by the DFG 
Sonderforschungsbereich/Transregio SFB/TR9.
\vspace{-0.625cm}

\bibliography{Literature-cond-nf2.bib}

\end{document}